\documentclass[11pt,aps,prb,superscriptaddress,showpacs]{revtex4}
\usepackage[dvips]{graphicx}
\usepackage{color}
\usepackage{bm}
\usepackage{epsfig}
\usepackage{amsmath,amssymb,amscd}
\begin{document}
\title{
Structure of energy level degeneracy of a single-spin model
from a view point of symmetry of the spin anisotropy and its
nontrivial spin($S$)-dependence on the higher order anisotropy
}
\author{Keigo Hijii}
\author{Seiji Miyashita}
\affiliation{
Department of Physics, Graduate School of Science,
University of Tokyo, Bunkyo-ku, Tokyo 113-0033, Japan
}
\affiliation{
CREST, JST, 4-1-8 Honcho Kawaguchi, Saitama, 332-0012, Japan
}
\begin{abstract}
We study structure of the gapless points (diabolical points)
at zero magnetic field ($H_z=0$) of single-spin models with spin anisotropies.
Nontrivial appearance of diabolical points at finite transverse field $H_x$
has been studied from the view point of interference of the Berry phase,
and related phenomena have been experimentally found
in the single molecular magnet Fe$_8$.
We study effects of the orthorhombic single-ion anisotropy
$E(S_+^2+S_-^2)$ and find a symmetry associated with the degeneracy,
which provides a clear picture of the global structure of energy
level diagram including the excited states.
Moreover, we study effects of the higher order anisotropy $C(S_+^4+S_-^4)$,
and find that, in contrast to the semiclassical limit $(S\rightarrow\infty)$,
location of a pair annihilation of the diabolical point
does not coincides with a point at which a pair of diabolical points
appears in nonzero $H_y$ space(bifurcation points).
Distance between the annihilation and bifurcation
points vanishes when $S\rightarrow\infty$,
which restores the semiclassical result.
We obtain a complete structure of the diabolical points
in the $(C,H_x)$ plane.
\end{abstract}



\maketitle


\section{Introduction}
\hspace*{\parindent}
Single molecular magnets, e.g., Mn$_{\rm 12}$,Fe$_{\rm 8}$ and V$_{\rm 15}$,
are interesting objects from both of theoretical and experimental points
of view in physics and chemistry
\cite{GSV2006,Wernsdorfer2001,GS2003,BP2004,BKHNDRHCC2005}.
Because those molecules consist of small number of magnetic atoms,
the energy levels are discrete.
There, we observe characteristics due to
quantum mechanical motion of the wave function.
In particular, in the high spin molecular magnets with an easy-axis anisotropy,
such as Mn$_{\rm 12}$ and Fe$_{\rm 8}$,
a step-like magnetization process where $M_z$ suddenly changes
has been observed in a sweep of the magnetic field.
This phenomenon is understood to be attribute to the quantum tunneling
between two values of $M_z$, and is called resonant tunneling.
\cite{BTLCS1999,TLBGSB1996,FSTZ1996,SOPSG1997,PBHHD1998,KGKTA2002}.
The energy level diagram as a function of the magnetic field $H_z$
consists of linear lines denoting the Zeeman energy (diabatic state).
At the crossing point of the energy levels, however,
they form an avoided level-crossing structure
due to some quantum mixing interactions
which cause nonzero matrix element between the crossing states.
When the field crosses these points,
the state undergoes adiabatic and nonadiabatic transitions.
This quantum mechanical aspect of magnetization process has been studied
from the view point of Landau-Zener-Stueckelberg mechanism
\cite{Landau1932,Zener1932,Stueckelberg1932,Miyashita1995}.
There,
the energy gap and sweeping velocity determine properties of the transition.
By making use of this formula, determinations of the energy gaps
have been performed\cite{WS1999,UMK2002,RAMKL2005}.
Besides the high spin molecules,
there have been also found various types of magnetization processes
which reflect the quantum mechanical aspects of
specific energy level diagram of the systems.
\cite{SMD1999,CWMBB2000,SM2001,CMNKHSRD2006,BGMTMB2008}.
These systems have attracted attentions also from view points of
possible applications, for example,
a basic component of a quantum computer\cite{LD2001}.

The energy gap is understood as a tunnel splitting of the energy levels.
That is,
by tunneling between classically degenerate minima of a potential,
the degeneracy is broken.
The idea of quantum tunneling of magnetization was proposed
by Bean and Livington\cite{BL1959},
and the first theoretical description was given by
Chudnovsky\cite{Chudnovsky1979}.
This tunneling phenomenon can be characterized by the instanton solution
in the semiclassical treatments\cite{WH1983,Coleman1985,Rajaraman1987}.
Thus, usually the ground state in finite quantum systems is unique.

However, in some situation, a degeneracy can exist as has been predicted by
Bogachek and Krive\cite{BK1992}.
The point at which the energy gap vanishes is called a "diabolical point"
\cite{BW1984}.
It was pointed out that an interference of Berry phase\cite{Berry1984}
plays an important role in small magnetic particles
\cite{LDG1992,DH1992,CD1993}.
Garg studied this phenomenon by
studying destructive interference of the Berry phase
by using the spin coherent state path integral formulation.
He showed that the tunnel splitting at $H_z=0$ is quenched
in a single spin system of a large spin $S$
with biaxial anisotropy of the terms $(-DS_z^2+E(S_+^2+S_-^2))$
under nonzero transverse fields $H_x$\cite{Garg1993,GKPS2003}
even when Kramers' theorem is inapplicable.
There, the tunnel splitting is found to oscillate as a function of
the transverse field.
That is, energy gaps vanish at some values of
the transverse field $H_x$.
Villain and Fort studied a case of large spin in a weak external field limit
\cite{VF2000}.
They rederived Garg's result, and extended the study in the ($H_x,H_z$) plane.
Ke\c{c}ecio\u{g}lu and Garg obtained exact locations of diabolical points
algebraically in a model Hamiltonian\cite{KG2001}.

Werensdorfer and Sessoli experimentally observed
the oscillating behavior of tunnel splitting in the molecular magnet
[Fe$_8$O$_2$(OH)$_{12}$(tacn)$_{6}$]$^{8+}$ (called Fe$_8$)\cite{WS1999}.
This spin system consists of eight Fe atoms each of which has $S=5/2$
conforming a ferrimagnetic structure.
The ground state of this molecule has the total spin $S=10$
\cite{BDGSS1996}.
This material is well described by a single large spin model.
They measured tunnel splitting of this material using
the Landau-Zener-Stueckelberg theory.
There, it is found that the number of diabolical points
is smaller than that expected from $S$,
which is called ``the missing paradox''.

Effects of the higher order anisotropy $C(S_+^4+S_-^4)$ are also studied.
Ke\c{c}ecio\u{g}lu and Garg explained the missing paradox
as an effect of the higher order anisotropy\cite{KG2002,KG2003}.
Bruno pointed out a pair annihilation of diabolical points in
the $(C,H_x)$ plane and they move to the nonzero $H_y$ space
\cite{Bruno2006}.
They discussed the case with the large $S$ limit
using spin coherent state path integral formulation.

In the present paper, we point out that
the mechanism of degeneracy
at finite values of $H_x$ can be understood
from a view of a kind of parity effect in the eigenvalues of $S_x$
which is directly obtained from the symmetry of the Hamiltonian
of the system.
This symmetry argument provides a clear picture of the global structure
of energy level diagram including the excited states.

Moreover, we study effects of the higher order anisotropy
$C(S_+^4+S_-^4)$ on positions of diabolical points in the $(C,H_x)$ plane,
and determine a complete structure of diabolical points in the plane.
There, we find three types of pair annihilation of the diabolical points,
and also find out to where the diabolical points move from the plane.
It should be noted that, in the case of finite $S$,
the pair annihilation point at finite $H_x$ does not coincide with
the point where a pair of diabolical points appears in nonzero $H_y$
space (bifurcation point)
in contrast to the semiclassical case ($S\rightarrow\infty$)
\cite{Bruno2006}.
We find that the distance between the annihilation and bifurcation
points vanishes when $S\rightarrow\infty$,
Namely, the semiclassical result is restored in this limit.
We also study a difference in the structure of diabolical points
for odd and even values of $S$, which should be related to the parity effect
pointed in the literature\cite{BTLCS1999}.

This paper is organized as follows.
In Sec.2, we introduce a single spin model of single molecular magnets.
In Sec.3, we study symmetry of the Hamiltonian of the single-spin model
in relation to the nontrivial degeneracy.
In Sec.4, we discuss effects of the higher order anisotropy.
Finally, in Sec.5, we summarize the present results.

\section{Model}
\hspace*{\parindent}
In this paper, we study structures of energy level diagram of
a large spin model described by
\begin{equation}
{\mathcal H} = - D S_z^2 + E \left( S_+^2 + S_-^2 \right)
+ C \left( S_+^4 + S_-^4 \right)
-{\bm H}\cdot{\bm S},
\label{eq:Hamiltonian0}
\end{equation}
where ${\bm S}$ is a spin operator with three component ($S_x,S_y,S_z$),
${\bm H}$ is an external magnetic field ($H_x,H_y,H_z$).
The terms of $D,E$ and $C$ represent the single-ion anisotropies.
When $D$ and $E(<D)$ are positive,
the easiest axis is the $z-$axis $(-DS_z^2)$,
and the hardest axis is the $x-$axis $(2ES_x^2)$.
This large spin model is used
to study properties of single molecular magnets
such as Mn$_{\rm12}$ and Fe$_{\rm 8}$.
For these molecules, the total spin $S$ of the ground state can be
regarded to be $S=10$\cite{CGS1991,BDGSS1996}.

In particular, we study effects of the system parameters on the energy levels,
and discuss the behavior of the diabolical points,
at which the ground state is degenerate at $H_z=0$ as has
mentioned in Introduction.
Throughout the paper, we take $D$ as a unit of energy ($D=1$).

\section{Symmetry of the model with biaxial anisotropy
under an external field $H_x$}
\subsection{Special symmetric point}
\hspace*{\parindent}
As mentioned in Introduction,
the problem of the diabolical point has been studied extensively
for the model (\ref{eq:Hamiltonian0}).
There, the ground state degeneracy at $H_z=0$ is studied as a function
of $H_x$, and found that the energy gap disappears at certain values of
$H_x$.
Generally, disappearance of the gap is associated with existence of
a kind of symmetry.
So far, the symmetry of the model has been discussed in the
path-integral formulation,
where the gap disappearance is attributed to a destructive interference
of the Berry phase.

In this section, we study the symmetry of the model (\ref{eq:Hamiltonian0})
with $C=0$ and the magnetic field along $x$-axis:
\begin{equation}
{\mathcal H}= -D S_z^2 + E \left( S_+^2 + S_-^2 \right) - H_x S_x,
\label{eq:HamiltonianC0}
\end{equation}
from a view point of explicit form of the Hamiltonian
consisting of spin operators.

\begin{figure}[ht]
\begin{center}
\epsfig{file=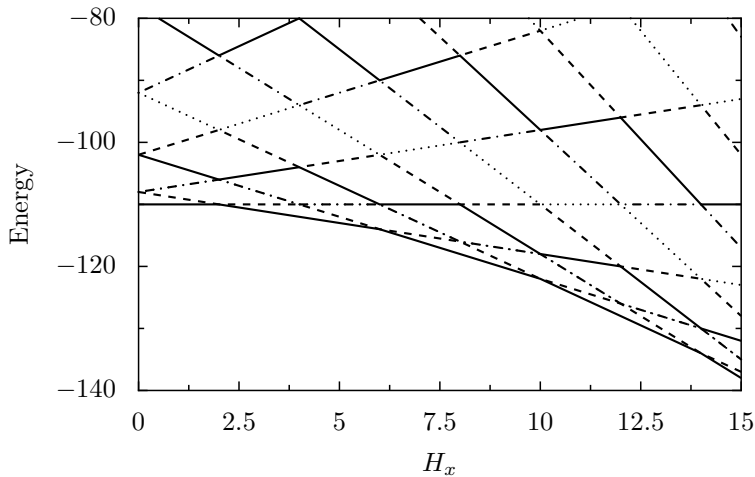,width=10cm}
\caption{Energy diagram of the low-lying levels of the system
(\ref{eq:Hamiltonian0}) with $S=10$
as a function of the field $H_x$ for $E=0.5$.
}
\label{fig:E050}
\end{center}
\end{figure}

\begin{figure}[ht]
\begin{center}
\epsfig{file=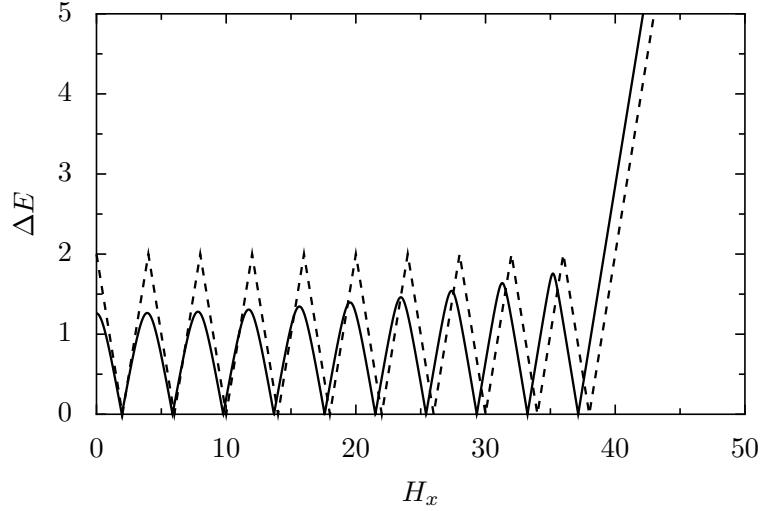,width=10cm}
\caption{
Energy gap between the lowest energy and the first excited energy
of the system Eq.(\ref{eq:HamiltonianC0}) with $S=10$
as a function of the transverse field $H_x$.
The solid line is the case of $E=0.485$,
the dashed line is the case of $E=0.5$.
}
\label{fig:sawtooth}
\end{center}
\end{figure}

Because we consider the case that the principal anisotropy axis is along
the $z$-axis,
naively we consider that the existence of $H_x$ destroys
the symmetry of the Hamiltonian.
However, it should be noted that at a certain combination of $D$ and
$E$, i.e.,
\begin{equation}
E=0.5D,
\end{equation}
the Hamiltonian can be expressed as follows
\begin{eqnarray}
{\mathcal H}_0 &=& -D S_z^2 + D \left( S_x^2 - S_y^2 \right) -H_x S_x
\nonumber \\
&=& 2DS_x^2 -H_x S_x -DS\left(S+1\right).
\end{eqnarray}
This Hamiltonian only consists of $S_x$,
and thus it is commutative with $S_x$.
Therefore, this Hamiltonian can be diagonalized simultaneously with $S_x$,
where the eigenstates are
\begin{equation}
S_x \left| M_x \right> = M_x \left| M_x \right>,
\quad M_x=-S,-S+1,\cdots,S.
\end{equation}
In this system,
the energy levels are linear as a function of $H_x$, and cross each other
without gap.
Because $D$ is positive,
at $H_x=0$ the ground state is a state of $M_x=0$,
i.e., $\left|M_x=0\right>$,
For $S=10$, the ground state energy is $-110D$.
The first excited state is degenerate and they have $M_x=\pm 1$.
When we increase $H_x$, the ground state is replaced by a state
with a larger magnetization $M_x+1$ sequentially.
That is, at $H_x=2$, the energy level of state $\left|M_x=1\right>$
crosses with that of $\left|M_x=0\right>$,
then $\left|M_x=1\right>$ becomes the ground state.
Similarly, the ground state magnetization changes to
$M_x=2,3,\cdots$ at $H_x=6,10,\cdots$, respectively.
In Fig.~\ref{fig:E050},
we depict the energy diagram of the model of Eq. (\ref{eq:HamiltonianC0})
as a function of the field $H_x$.
In Fig.~\ref{fig:sawtooth},
we plot the energy gap between the ground state energy ($E_{\rm G}$)
and the first excited energy ($E_1$)
\begin{equation}
\Delta E = E_1-E_{\rm G},
\end{equation}
by dashed lines as a function of $H_x$.
There, we see a saw-tooth shape as shown.

\subsection{General biaxial anisotropy}
\hspace*{\parindent}
Next, we consider the case with $E \ne 0.5D$.
We set
\begin{equation}
E=0.5D+\Delta.
\end{equation}
The Hamiltonian becomes
\begin{equation}
{\mathcal H}={\mathcal H}_0+{\mathcal H}',
\end{equation}
with
\begin{equation}
{\mathcal H}' = \Delta \left( S_+^2 + S_- ^2 \right)
= 2\Delta \left( S_x^2 - S_y^2 \right).
\end{equation}
Here, the states $\left|M_x\right>$ are no more the eigenstates
of the Hamiltonian ${\mathcal H}$.
The effects of the term $S_y^2$ is expressed in terms of
the raising ($S_x^+$) and lowering ($S_x^-$) operators
of for $M_x$ as
\begin{eqnarray}
S_y^2 &=& \left( \frac{1}{2} \left(  S_x^+ + S_x^- \right) \right)^2
\nonumber \\
&=& \frac{1}{4}\left(S_x^{+2}+S_x^+ S_x^-+S_x^- S_x^++S_x^{-2}\right).
\end{eqnarray}
This term causes the change of $M_x$ by two.
The explicit matrix element of this operator is
$\left< M_x=m \right|S_y^2\left| M_x=n \right>$
\begin{eqnarray}
&=&
\frac{1}{4}
\left( S \left( S + 1 \right) - n \left( n + 1 \right) \right)^{\frac{1}{2}}
\left( S \left( S + 1 \right) - \left( n +1 \right)
\left( n + 2 \right) \right)^{\frac{1}{2}}
\delta_{m,n+2}
\nonumber \\
 & & +
\frac{1}{4}
\left( 2S\left( S + 1 \right) - 2 n^2 \right)
\delta_{m,n}
\nonumber \\
 & & +
\frac{1}{4}
\left( S \left( S + 1 \right) - n \left( n - 1 \right) \right)^{\frac{1}{2}}
\left( S \left( S + 1 \right) - \left( n - 1 \right)
\left( n - 2 \right) \right)^{\frac{1}{2}}
\delta_{m,n-2}.
\end{eqnarray}
This term mixes the eigenstates $\left|M_x=m\right>$ and
$\left|M_x=n\right>$
when
\begin{equation}
\left|m-n\right|=2,
\end{equation}
and thus it opens a gap in the crossing points with even values of $|m-n|$
in the energy diagram in Fig.~\ref{fig:E050}.
In contrast,
it does not open a gap between
$\left| M_x = m\right>$ and $\left| M_x = m \pm 1 \right>$,
because
\begin{equation}
	\left< M_x=m \right| S_y^2 \left| M_x= m \pm 1 \right> =0,
\end{equation}
and,
\begin{equation}
	\left< M_x=m \right|
	S_y^2
	\left| M_x=n \right> \left< M_x=n \right|
	S_y^2
	\left| M_x=m \pm 1 \right>=0
\end{equation}
for all the possible integer values of $n$.

\begin{figure}[ht]
\begin{center}
\epsfig{file=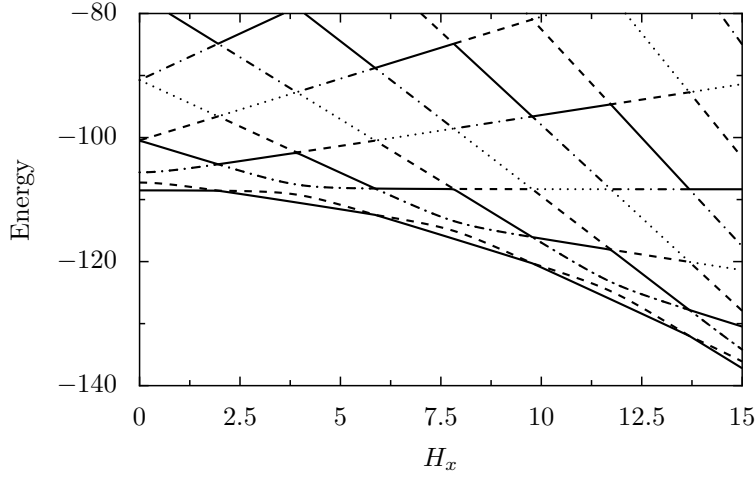,width=10cm}
\caption{
A ribbon like structure of energy diagram of the low-lying levels
of the system Eq.(\ref{eq:HamiltonianC0}) with $S=10$ and $E=0.485$
as a function of the field $H_x$.
}
\label{fig:E0485}
\end{center}
\end{figure}

\begin{figure}[ht]
\begin{center}
\epsfig{file=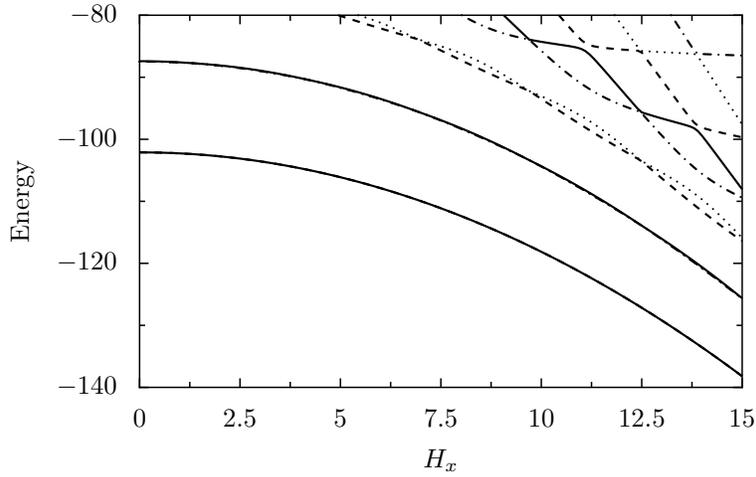,width=10cm}
\caption{Energy diagram of the low-lying levels of the system
Eq.(\ref{eq:HamiltonianC0})
with $S=10$ and $E=0.3$
as a function of the field $H_x$.
The lowest energy and the first excited energy almost degenerate
in this vertical axis scale.
The second excited energy and the third excited energy are also almost degenerate.
}
\label{fig:E030}
\end{center}
\end{figure}

\begin{figure}[ht]
\begin{center}
\epsfig{file=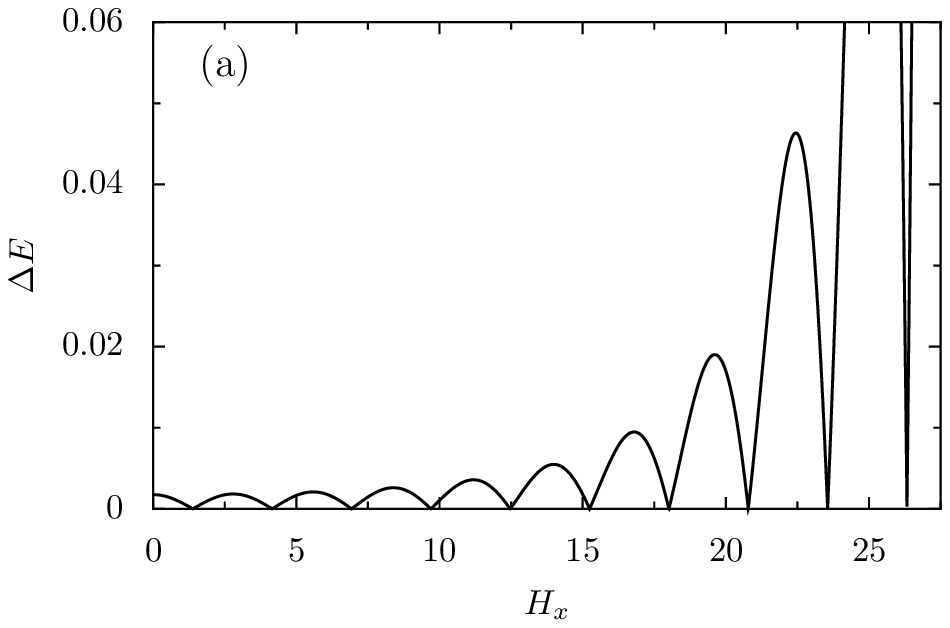,width=7cm}
\epsfig{file=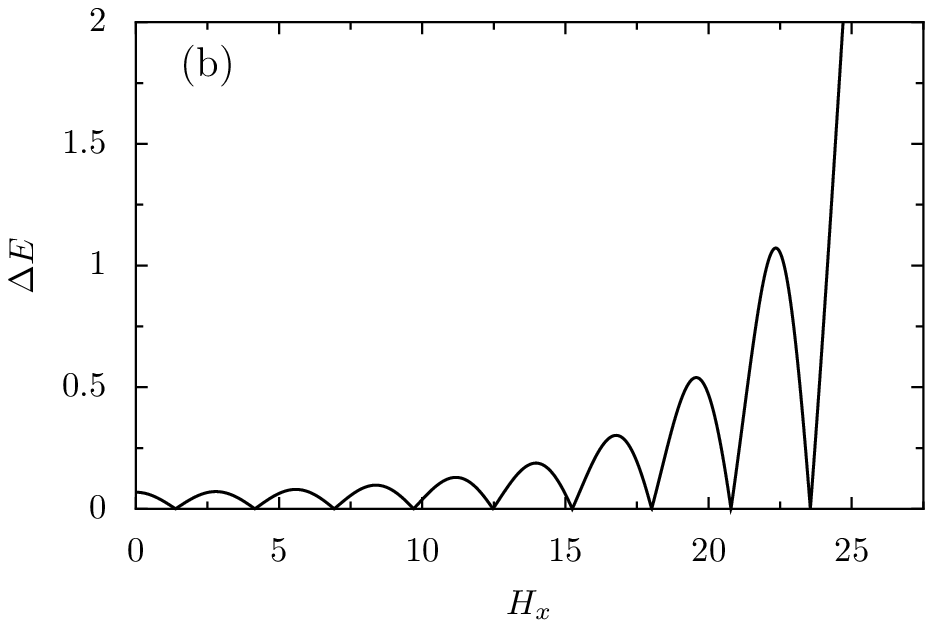,width=7cm}
\caption{
(a) Energy gap between the lowest energy and the first excited energy
for $E=0.3$ as a function of the field $H_x$.
There are 10 diabolical points.
(b) Energy gap between the second excited energy and the third excited energy
for $E=0.3$.
In this case, there are 9 diabolical points.
}
\label{fig:deltaE030}
\end{center}
\end{figure}

Therefore,
when the difference of the magnetization $M_x$
between the ground state and the first excited state is one,
the cross points in Fig.~\ref{fig:E050} remain gapless points $(\Delta E=0)$.
On the other hand, those of the difference two change to avoided level crossings.
By this effect of $S_y^2$, the energy diagram has a ribbon-like shape
as depicted in Fig.~\ref{fig:E0485}, and
the $H_x$ dependence of the gap is smoothed as depicted
in Fig.~\ref{fig:sawtooth} by a solid curve.
It should be noted that the value of $E/D$ is 0.082 for Fe$_8$ and
is much smaller for Mn$_{12}$.
Here we used a large value of $E/D$ just because of
the convenience for drawing the figure.
If we use a small value of $E/D$,
the energy difference is too small to see.
The physical mechanism is the same irrespective of the value, and
here we use a large value.
If we decrease the value of $E$ down to $E=0.3$,
the ground state and the first excited state almost degenerate
as depicted in Fig.~\ref{fig:E030}.
There, the energy gap $\Delta E$ has a shape
which has often appeared in literature
(Fig.~\ref{fig:deltaE030}).

\section{Effects of a higher order anisotropy}
\hspace*{\parindent}
In single molecular magnets with large spins,
e.g., Mn$_{12}$~\cite{MHCAGIC1999,HPDHSB1998,BGS1997}
and Fe$_8$~\cite{WS1999},
existence of the higher order anisotropic term
\begin{equation}
{\mathcal H}''= C\left( S_+^4 + S_-^4 \right)
\label{eq:4th_Hamiltonian}
\end{equation}
has been suggested.
In this section, we study effects of this fourth order anisotropy.
The Hamiltonian without the magnetic field is
\begin{equation}
{\mathcal H}=-DS_z^2+E\left(S_+^2+S_-^2\right)+{\mathcal H}''.
\label{eq:4th_Hamiltonian_all}
\end{equation}
Here it should be noted as follows.
Because $S_{+}=S_x+iS_y$ and $S_{-}=S_x-iS_y$, and
\begin{equation}
S_{+}^4 + S_{-}^4 =
2 S_x^4 + 2 S_y^4
- 6 S_x^2 S_y^2
- 6 S_y^2 S_x^2
-4i\left( S_x S_z S_y - S_y S_z S_x \right)
-2 S_z^2
\end{equation}
Thus, in the representation which diagonalize $M_x$ i.e.,
$\{ |M_x\rangle \}$, it is given by
\begin{align}
S_{+}^4 + S_{-}^4 =&
2 S_x^4 + \frac{1}{8} \left( S_x^+ + S_x^- \right)^4
- \frac{3}{2} S_x^2 \left( S_x^+ + S_x^- \right)^2
- \frac{3}{2} \left( S_x^+ + S_x^- \right)^2 S_x^2
\nonumber \\
& + S_x \left( S_x^+ - S_x^- \right)\left( S_x^+ + S_x^- \right)
-
\left( S_x^+ + S_x^- \right)\left( S_x^+ - S_x^- \right) S_x
\nonumber \\
& + \frac{1}{2} \left( S_x^+ - S_x^- \right)^2
\end{align}
which can change the value of $M_x$ by multiples of 2.

Therefore, nonzero components of matrix elements of the fourth term are
\begin{eqnarray}
\left< M_x=m \right|
{\mathcal H}''
\left| M_x=m \right>,
\nonumber  \\
\left< M_x=m \right|
{\mathcal H}''
\left| M_x=m \pm 2 \right>,
\end{eqnarray}
and
\begin{displaymath}
\left< M_x=m \right|
{\mathcal H}''
\left| M_x=m \pm 4 \right>.
\end{displaymath}
Because
\begin{equation}
\left< M_x=m \right|
{\mathcal H}''
\left| M_x=m \pm 1 \right>
=0,
\label{eq:4th_Hamiltonian_pm1}
\end{equation}
the fact that the gap opens only at crossing points
where the magnetization $M_x$ differs by two maintains.

\subsection{Dependence on $C$ at fixed $E$}
\begin{figure}[ht]
\begin{center}
\epsfig{file=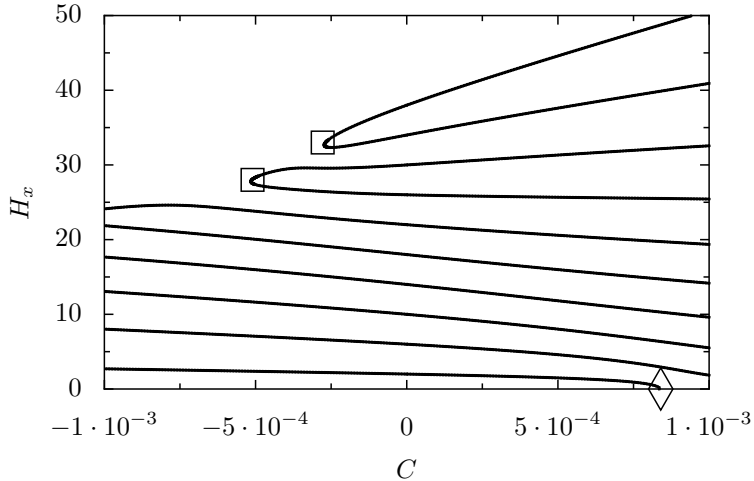,width=100mm}
\caption{
Diabolical points between the lowest energy level and the first excited energy
level on the $(C,H_x)$ plane for the $E=0.5$.
The symbol ($\square$) denotes the type I annihilation points.
The symbol ($\diamond$) denotes the type II annihilation points.
}
\label{fig:dp_E050}
\end{center}
\end{figure}

\hspace*{\parindent}
First let us study the behavior of the diabolical points on $C$
at fixed value of $E$.
We plot the change of the diabolical points in a coordinate $(C,H_x)$
in Fig.~\ref{fig:dp_E050}.
As far as $\left|C\right|$ is small,
the number of diabolical points is the same as that of $C=0$.
However, for large $\left|C\right|$ cases,
pairs of diabolical points disappear from the figure.
We call this point $(C,H_x)$ ``type I an annihilation point''
which is shown by ($\square$) in Fig.~\ref{fig:dp_E050}.
The pair annihilation occurs from the side of large $H_x$
when $C$ decreases in the negative $C$ region.
In the positive side, diabolical points are drawn into the $H_x$ axis
sequentially.
At the $H_x$ axis, the diabolical point combines with that from
the negative $H_x$ side, and disappears,
which we call ``Type II annihilation points'', and denote it by ($\diamond$)
in Fig.~\ref{fig:dp_E050}.
At these annihilation points, the diabolical points move to a nonzero $H_y$ region.

\begin{figure}[ht]
\begin{center}
\epsfig{file=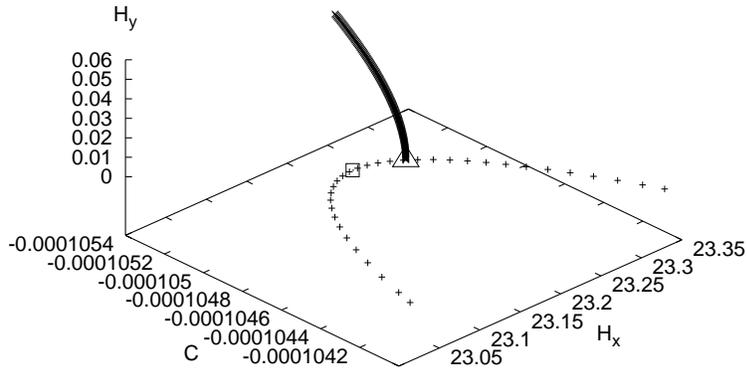,width=10cm}
\caption{
The branch of diabolical points between the lowest energy and first
excited energy with largest $H_x$
in the case of $S=10$  and $E=0.3$.
The symbol ($\triangle$) denotes the bifurcation point.
The symbol ($\Box$) denotes the annihilation point.
}
\label{fig:dp_S10_space}
\end{center}
\end{figure}

First, we show the motion of diabolical points around the type I
annihilation point.
In Fig.~\ref{fig:dp_S10_space},
we plot the motion of diabolical points in the largest $H_x$ values
in a $H_x>0$ subspace.
There, we find that a pair of diabolical points is created
in nonzero $H_y$ region at a point.
We denote this point by the symbol ($\triangle$).
We call this point ``a bifurcation point''.
Here, it should be noted that the point of the creation of the pair is not
the point of the annihilation of the pair on the $(C,H_x)$ plane.
\begin{figure}[ht]
\begin{center}
\epsfig{file=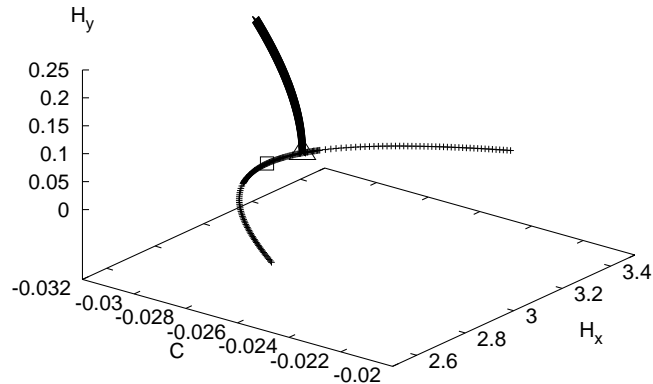,width=10cm}
\caption{
The branch of diabolical points between the lowest energy and first
excited energy with largest $H_x$
in the case of $S=2$ and $E=0.3$.
The symbol ($\triangle$) denotes the bifurcation point.
The symbol ($\Box$) denotes the annihilation point.
}
\label{fig:dp_S2_space}
\end{center}
\end{figure}
We find that this separation of the annihilation point and the bifurcation
point exists in all the finite values of $S$.
In Fig.~\ref{fig:dp_S2_space}, we show the case of $S=2$,
where we find the same type of structure.
The separation is much larger than the case of $S=10$.

The effect of the fourth order anisotropy has been discussed by Bruno
\cite{Bruno2006}.
His argument is the following.
There is a critical value of $C=C_{\rm c}$
where two diabolical points collide,
and at this point the bifurcation takes place.
That is,
a pair of two diabolical points appears at the type I annihilation point.
However, we find that
the bifurcation point is different from the annihilation point,
and appears at a larger value (smaller $\left|C\right|$) of $C$.
This means that the number of diabolical points are not preserved
on the $(H_x,H_y)$ plane
when we change $C$.
This fact is different from Bruno's argument.
In his arguments,
the number of diabolical points on the $(H_x,H_y)$ plane is preserved
except at $C_{\rm c}$
On the other hand,
our numerical result shows that the number of diabolical points
on the $(H_x,H_y)$ plane can change with the value of $C$.
Bruno's discussion is based on the large $S$ limit.
Thus, we study $S$ dependence of the separation of the annihilation
and bifurcation points.

\begin{figure}[ht]
\begin{center}
\epsfig{file=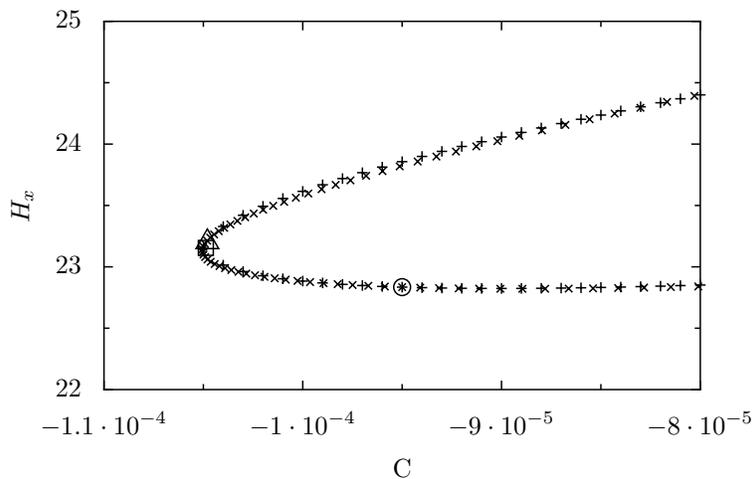,width=10cm}
\caption{
Fitting of diabolical points using a rotated parabola function.
The symbols ($+$) denote the bare numerical results for the diabolical points.
The symbol ($\triangle$) denotes the bifurcation point.
The symbol ($\Box$) denotes the annihilation point.
The symbols ($\times$) denote points on a rotated parabola obtained by fitting,
and the symbol ($\bigcirc$) denotes the origin of the fitted parabola.
(Because the scales of axes of $H_x$ and $C$ are different,
the point denoted by the circle does not look like the origin.)
}
\label{fig:parabola}
\end{center}
\end{figure}

Here, we investigate structure of the diabolical points near annihilation points.
In Fig.~\ref{fig:dp_E050},
a pair of diabolical points near annihilation points has
a parabola-like structure on the $(C,H_x)$ plane.
Thus, we try to fit the curve using a rotated parabola function
$(a^2C^2+2abH_xC+b^2H_x^2+cC+dH_x+e=0)$ with constants $(a,b,c,d,e)$.
The fitting is given in Fig~\ref{fig:parabola}.
The origin of this rotated parabola where the diabolical point
is located at $(C,H_x)\sim (-0.000095,22.83468)$,
which is indicated ($\bigcirc$).
The point is not the annihilation point,
and it is not the bifurcation point neither.
This fact is indicates that the bifurcation does not occur
at the origin of the parabola which is a special point
of this figure.

Now, we study $S$-dependence of the distance
between the bifurcation point and the annihilation point.
We define two quantities,
\begin{equation}
\Delta C \equiv C_{\rm bif} - C_{\rm ann},
\end{equation}
and
\begin{equation}
\Delta H_x \equiv H^{x}_{\rm bif} - H^x_{\rm ann},
\end{equation}
where $C_{\rm bif}$ and $H^x_{\rm bif}$ are values of
bifurcation points,
and $C_{\rm ann}$ and $H^x_{\rm ann}$ are values of
annihilation points.

\begin{figure}[ht]
\begin{center}
\epsfig{file=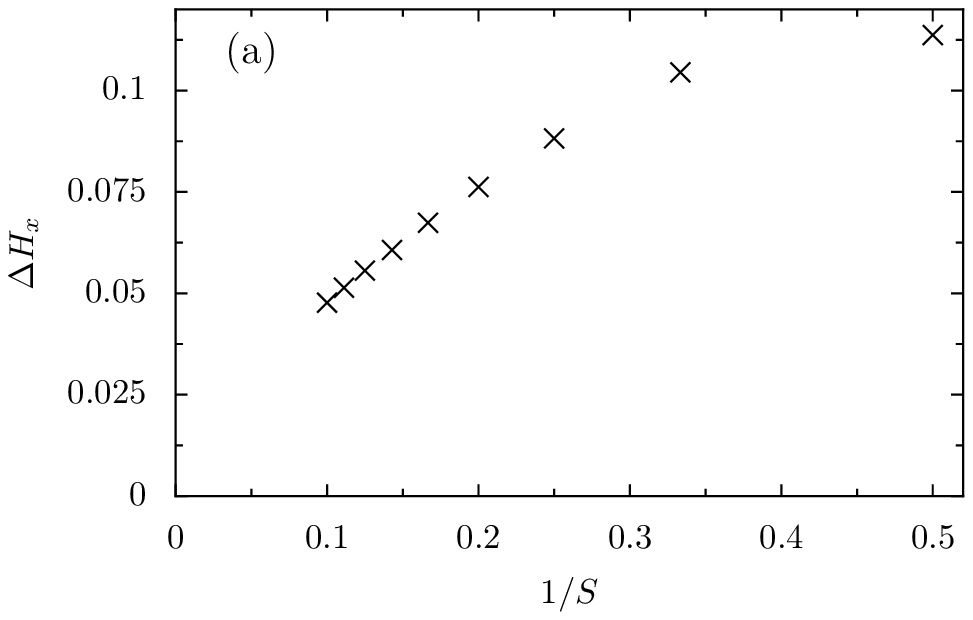,width=7cm}
\epsfig{file=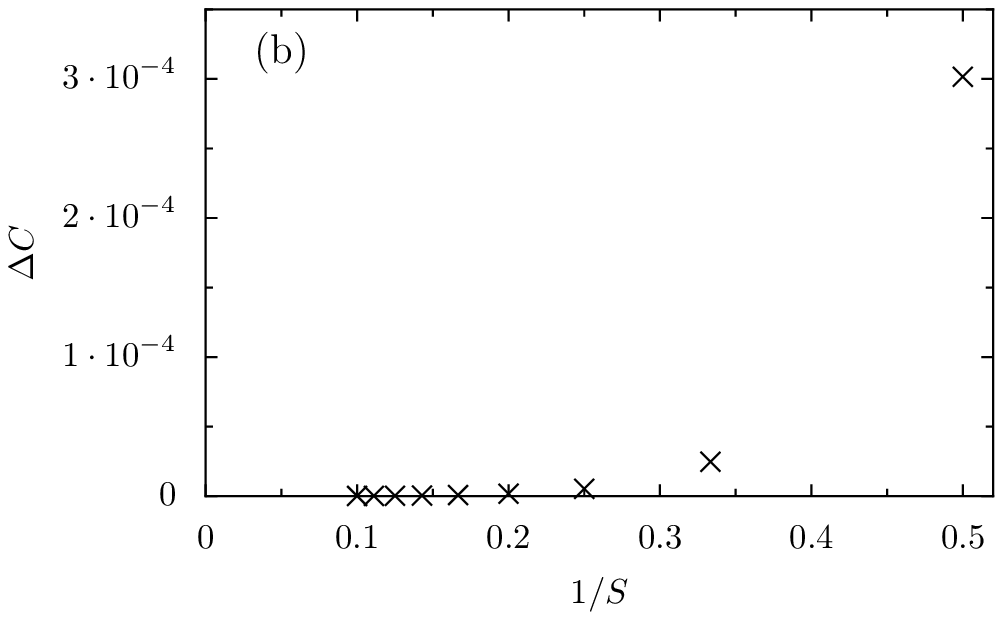,width=7cm}
\caption{
(a):$\Delta H_x$ as a function of $1/S$ with $E=0.3$.
(b)$\Delta C$ as a function of $1/S$ with $E=0.3$.
}
\label{fig:del_H_C}
\end{center}
\end{figure}

We plot
$\Delta C$ and $\Delta H_x$ as a function of $1/S$,
in Fig.~\ref{fig:del_H_C}.
In these figures, we find that both $\Delta C$ and $\Delta H_x$
rapidly decrease,
when we increase $S$.
Thus, our numerical results are consistent with Bruno's arguments
in the large $S$ limit.
But, it should be noted that at finite values of $S$ the bifurcation point
and the annihilation point do not coincide,
which indicates there exists a nontrivial quantum effect.

\begin{figure}[ht]
\begin{center}
\epsfig{file=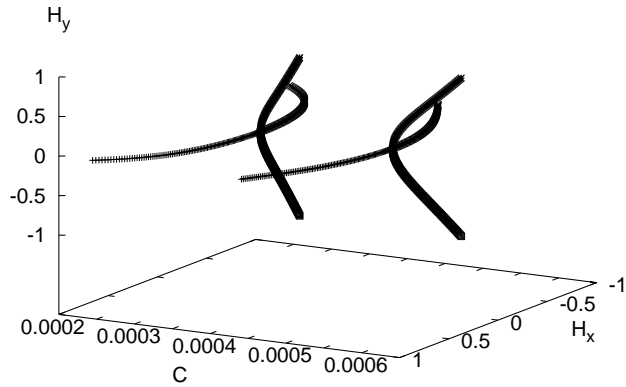,width=100mm}
\caption{
Diabolical points between the lowest energy level and the first excited energy
level on the $(H_x,C,H_y)$ space in the case of $E=0.3$.
}
\label{fig:dp_bif_C_positive}
\end{center}
\end{figure}

Next, we show the motion of the diabolical points around the type II
annihilation points.
There, two diabolical points move from $(C,H_x,H_y=0)$
to $(C,H_x,H_y\neq 0)$.
In Fig.~\ref{fig:dp_bif_C_positive},
we show this motion of diabolical points in the $(C,H_x,H_y)$ space.

\begin{figure}[ht]
\begin{center}
\epsfig{file=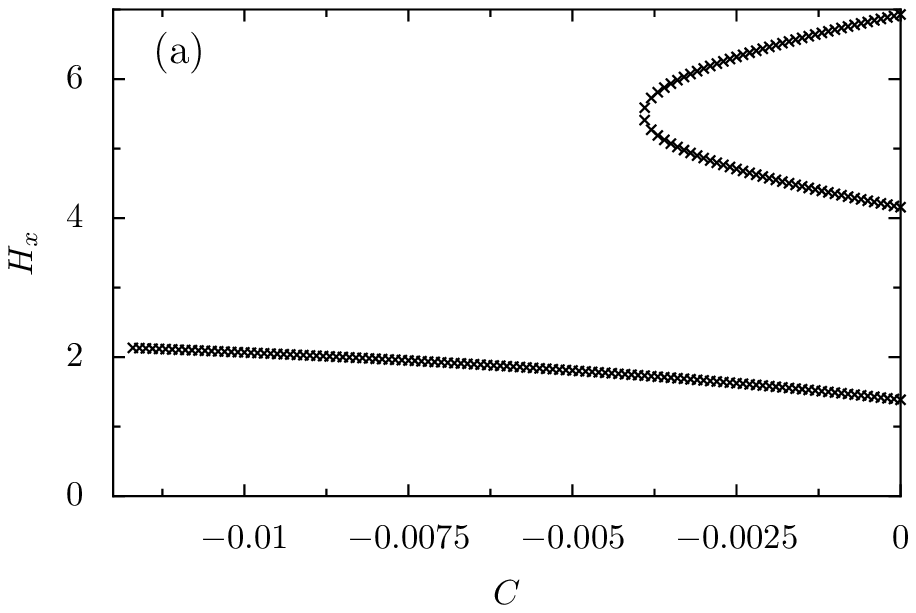,width=80mm}
\epsfig{file=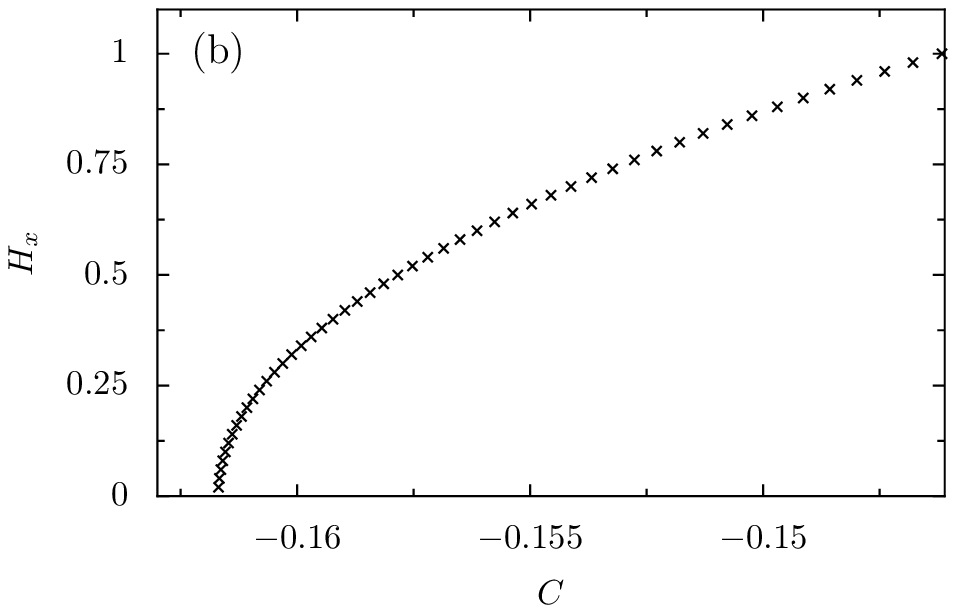,width=80mm}
\caption{
Behavior of diabolical points on the $(H_x,C)$ plane with $E=0.3$
for $S=3$ case:
(a) around the last pair annihilates.
and 
(b) the last one merges to the $C$ axis ($H_x=0$).
}
\label{fig:dp_fixed_e_c}
\end{center}
\end{figure}

\begin{figure}[ht]
\begin{center}
\epsfig{file=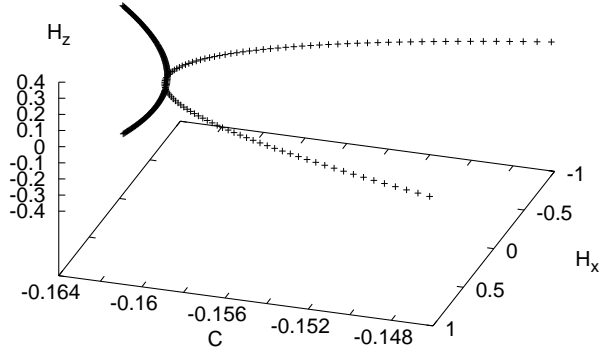,width=10cm}
\caption{
Diabolical points on $(H_x,C,H_z)$ space with $E=0.3$.
for $S=3$ case.
}
\label{fig:dp_fixed_e}
\end{center}
\end{figure}

As we saw above, the diabolical points disappear from the $(C,H_x)$ plane
by the pair annihilation.
In the case that $S$ is an odd integer,
there is an odd number of diabolical points in the $H_x(>0)$ region
of the $(C,H_x)$ plane.
There, the last one does not have a partner.
We study how the last point behaves in the $(C,H_x)$ plane.
In Fig.~\ref{fig:dp_fixed_e_c},
we show behavior of diabolical points of the model of $S=3$
in the $(C,H_x)$ plane.
In this case, there are three diabolical points in the region of $H_x>0$.
In Fig.~\ref{fig:dp_fixed_e_c}(a), we find the pair annihilates
around $C\sim -0.0039$.
There, the $H_x$ value of the last point increases
when $C$ decreases.
However, when $C$ decreases further,
it goes down and finally it merges to the $C$ axis
as shown in Fig.~\ref{fig:dp_fixed_e_c}(b),
and merges with the partner coming from the $H_x<0$ region.
We call this point ``the type III annihilation point''.
Interestingly in this case the diabolical points move to a nonzero
$H_z$ region
$(C,H_x(=0),H_y(=0),H_z(\neq 0))$.
but not a nonzero $H_y$ region
$(C,H_x(\neq 0),H_y(\neq 0),H_z(=0))$
as in the other cases.
We depict this behavior of diabolical points
in Fig.~\ref{fig:dp_fixed_e}.

In this way, all the diabolical points disappear from the $(C,H_x)$ plane
when $\left|C\right|$ becomes large,
and found three types of annihilation points.
By the above studies,
we figured out complete structure of diabolical points
in the $(C,H_x)$ plane.

\subsection{Dependence on $E$ at fixed $C$}
\begin{figure}[ht]
\begin{center}
\epsfig{file=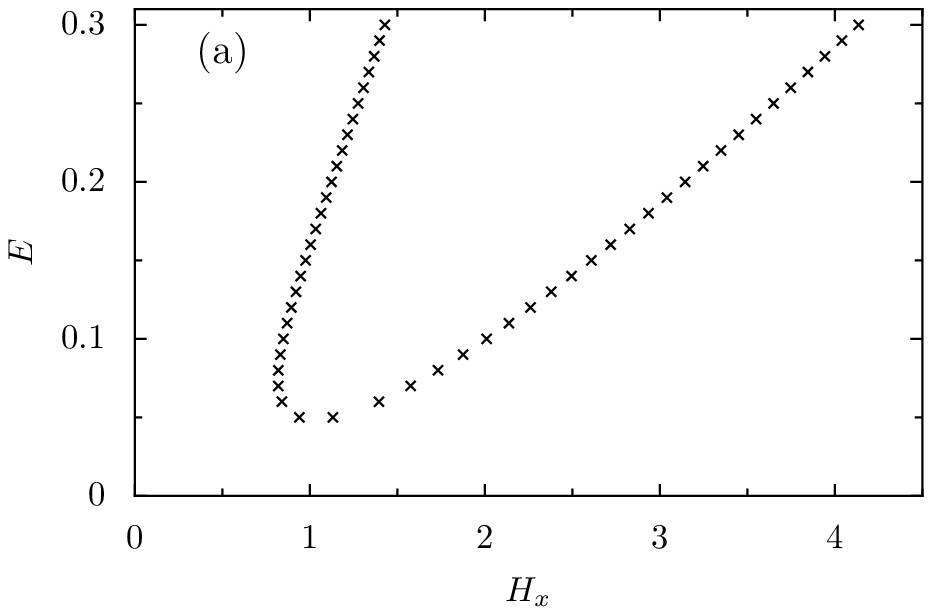,width=7cm}
\epsfig{file=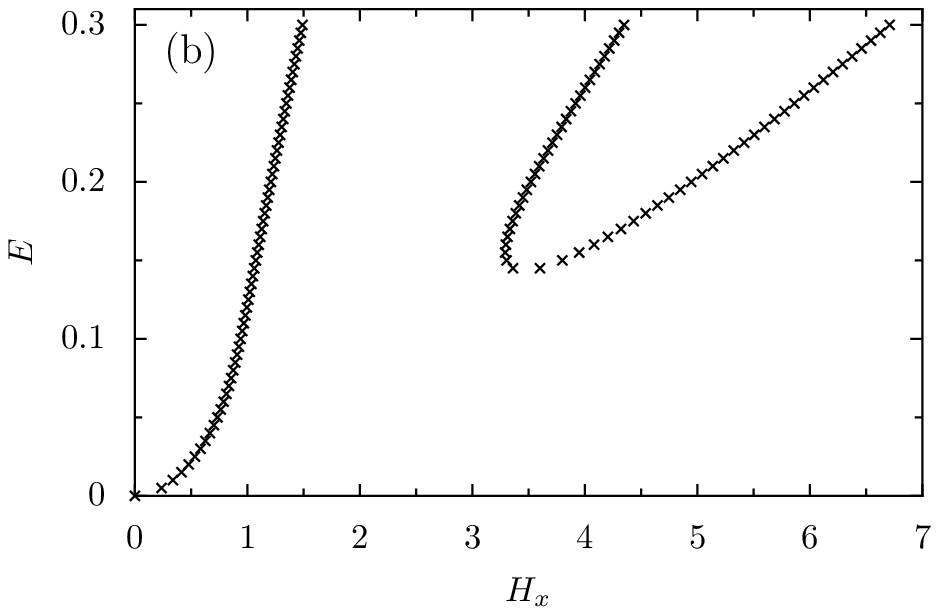,width=7cm}
\caption{
Diabolical points on the $(H_x,E)$ plane with $C=-0.001$.
(a) $S=2$, and (b) $S=3$.
}
\label{fig:dp_even_odd}
\end{center}
\end{figure}

\hspace*{\parindent}
So far, we studied the behavior in the $(C,H_x)$ plane.
Here let us study $E$ dependence of the diabolical points.
In Fig.~\ref{fig:dp_even_odd},
we show diabolical points on the $(H_x,E)$ plane for a fixed $C(=-0.001)$.
In Fig.~\ref{fig:dp_even_odd}(a), we show the case of $S=2$,
where the two diabolical points combine and annihilate
when $E$ becomes small.
This is a type I annihilation point.
There, they move to nonzero $H_y$ region.
In the case of $S=3$ cases, the last one diabolical point
moves to the origin $(H_x,E)=(0,0)$ as depicted in
Fig.~\ref{fig:dp_even_odd}(b).
This is a special case of the type III annihilation point.

The same type behavior is found in larger spin cases $(S=4,5,,,)$
(not shown).
This observation indicates that the ground state for $E=0$ is
two fold degenerate in the odd spin cases.
This is a degeneracy not related to Kramer's degeneracy,
because $S$ is integer.
We can easily understand this degeneracy.
For $E=0$, a Hamiltonian is described by
\begin{equation}
{\mathcal H}=-DS_z^2 + C \left( S_+^4 + S_-^4\right).
\end{equation}
If we set $C=0$,
$\left|M_z=-S\right>$ and $\left|M_z=S\right>$
give the two fold degenerate ground state,
where $S_z\left|M_z\right>=M_z \left|M_z\right>$.
For even spin cases, matrix element between the states
$\left|M_z=\pm S\right>$ is nonzero
\begin{equation}
\left< M_z=S \right|
\left( S_+^4 + S_-^4\right)^n
\left| M_z=-S \right>
\neq  0,
\end{equation}
because the difference of the magnetization $M_z$ (=$2S$)
is a multiple of 4,
where $n$ is an arbitrary integer.
On the other hand, and for odd spin cases,
the difference $2S$ is not a multiple of 4.
Thus,
\begin{equation}
\left< M_z=S \right|
\left( S_+^4 + S_-^4\right)^n
\left| M_z=-S \right>
=0.
\end{equation}
Therefore, quantum tunneling between the two states
does not occur,
and the ground state is two fold degenerate
in odd spin models for $E=0$ and $C\neq 0$ cases.

\section{Summary}
\hspace*{\parindent}
We investigated nontrivial degeneracy of eigenenergies
of single molecular magnets using the large single spin model.
In the parameter space $(E,C,H_x,H_y,H_z)$,
positions of the points at which the eigenenergies are degenerate
(diabolical points) are studied.
As has been pointed out, the model (\ref{eq:Hamiltonian0})
has diabolical points at nonzero $H_x$.
This fact seems nontrivial and has been studied in terms of the Berry phase
in the path-integral formulation\cite{Garg1993}.
We pointed out that the existence of diabolical points
at nonzero $H_x$ is understood from a view point of the
parity effect of the magnetization  in the  $x$ direction.

We also studied effects of the higher order anisotropy $C$.
For a small value of $\left|C\right|$, there are $S$ diabolical points
with positive values of $H_x$.
We studied behavior of those points when $\left|C\right|$ increases.
They move out from the $(C,H_x)$ plane by pair annihilations.
We found three types of annihilations.
In the positive $C$ case, each diabolical point moves to the $C$ axis,
and at the $C$ axis it combines with the partner coming from negative
$H_x$ region and they move to the nonzero $H_y$ region.
In the negative $C$ case, the diabolical points make a pair with neighbors
in the positive $H_x$ region.
We also found a pair creation of diabolical points in the nonzero $H_y$
region.
We should make emphasis that the annihilation points do not coincide
with the creation (bifurcation) points for finite values of $S$.
This is contrast to the case of $S\rightarrow\infty$,
which was studied by Bruno\cite{Bruno2006}.
The asymptotic behavior in the limit $S\rightarrow \infty$ was studied
and we found the distance between the annihilation and the bifurcation
points decreases to zero when $S$ increases.
Thus, the argument of semiclassical picture is valid,
but there exists an intrinsic quantum effect.
In the case of odd integer $S$,
one diabolical point remains unpaired and it moves to the $C$ axis
and make pair with a partner coming from negative $H_x$.
In this case, we found that they move to the nonzero $H_z$ region.

\section*{acknowledgement}
\hspace*{\parindent}
The authors thank to Keiji Saito for fruitful discussion.
The present work was supported by Grant-in-Aid for Scientific Research
on Priority Areas, and also and the Next Generation Super Computer
Project, Nanoscience Program from MEXT of Japan.
The numerical calculations were supported by the supercomputer center of
ISSP of Tokyo University.

\end{document}